\documentclass[conference]{IEEEtran}
\IEEEoverridecommandlockouts
\usepackage{cite}
\usepackage{amsmath,amssymb,amsfonts}
\usepackage{algorithmic}
\usepackage{graphicx}
\usepackage{textcomp}
\usepackage{xcolor}

\usepackage{mathtools}
\usepackage[cmintegrals]{newtxmath}

\renewcommand \Re{\operatorname{Re}}  

\def\BibTeX{{\rm B\kern-.05em{\sc i\kern-.025em b}\kern-.08em
    T\kern-.1667em\lower.7ex\hbox{E}\kern-.125emX}}

\begin{document}

\title{Improved OFDM Signal Cancellation through Window Estimation}

\author{\IEEEauthorblockN{Daniel Chew\IEEEauthorrefmark{1},
        Samuel Berhanu\IEEEauthorrefmark{1}, 
		Chris Baumgart\IEEEauthorrefmark{1}, 
		and A. Brinton Cooper\IEEEauthorrefmark{2}}
	\IEEEauthorblockA{\IEEEauthorrefmark{1}Applied Physics Laboratory, Johns Hopkins University\\}
	\IEEEauthorblockA{\IEEEauthorrefmark{2}Electrical and Computer Engineering, Johns Hopkins University}
}

\maketitle

\begin{abstract}
The ability to cancel an OFDM signal is important to many wireless communication systems including Power-Domain Non-orthogonal Multiple Access (PD-NOMA), Rate-Splitting Multiple Access (RSMA), and spectrum underlay for dynamic spectrum access. In this paper, we show that estimating the windowing applied at the transmitter is important to that cancellation. Windowing at the transmitter is a popular means to control the bandwidth of an Orthogonal Frequency Division Multiplexed (OFDM) symbol and is overlooked in most literature on OFDM signal cancellation. We show the limitation to the amount of cancellation that can be achieved without knowledge of OFDM windowing. We show that the window can be estimated from received samples alone, and that window estimate can be used to improve the signal cancellation. The window is estimated in the presence of noise and imperfect estimates of the center frequency offset (CFO) and the channel. We conclude with results using synthetic and over-the-air data where we demonstrate a 5.3 dB improvement to OFDM signal cancellation over existing methods in an over-the-air experiment. 
\end{abstract}


\section{Introduction}
Multi-User Interference Cancellation (MUIC) enables wireless users to address interference caused by multiple users accessing a single spectrum resource concurrently. MUIC is the process of modeling and reproducing a signal from one particular user for the purposes of removing that signal from the summed ensemble of all received signals. Successive Interference Cancellation (SIC) \cite{Miridakis2013} is a category of MUIC implementations in which cancellation is applied sequentially over multiple users in the ensemble. MUIC schemes have been proposed for numerous wireless systems, for example, MUIC and PD-NOMA have been proposed for 5G systems \cite{Dai2015} \cite{Ding2017}. The ability to cancel a signal in order to receive another is important for spectrum underlay, a dynamic spectrum access technique where the secondary users transmit \emph{under} the incumbent signal \cite{IEEE1901def}. SIC is also used in RSMA \cite{Clerckx2016} which is a multiple access scheme that has been projected to outperform PD-NOMA \cite{Mao2018}. As a final example, the work in both \cite{WiFiNOMATestbed2018} and \cite{WiFiNOMAAutomation2020} augment 802.11 with power-domain multiplexing schemes and SIC.    

The use of MUIC requires a signal model that includes the transmitted waveform and the impairments to be estimated. Several signal models appear in literature \cite{Manglayev2018} \cite{Miridakis2013} \cite{Clerckx2020RSMA1} \cite{Panayirci2010}. All of these describe the cancellation process as demodulation, remodulation, and then augmenting the remodulator output with estimated impairments. This process is described in \cite{Miridakis2013} but it does not specify which impairments are modeled. The signal model in \cite{Manglayev2018} includes signal amplitude and phase. The channel coefficients are estimated for the signal model in \cite{Clerckx2020RSMA1}. The work in \cite{Panayirci2010} employs estimates for channel coefficients and ``inter-channel interference" (ICI) meaning wireless impairments have caused the subcarriers of the OFDM symbol to no longer be orthogonal. 

These existing signal models may not account for enough parameters to achieve the desired level of cancellation. One often-overlooked signal-model parameter that can increase cancellation is OFDM windowing. Windowing an OFDM signal at the transmitter is a popular means to control the bandwidth of the signal and meet the spectral mask imposed by a given wireless standard \cite{Farhang2011}. An example can be found in the IEEE 802.11 standard which suggests using windowing but does not mandate it \cite{80211Standard2020}. The use, shape, and length of the window is left to individual vendors. Typically, the cyclic prefix is extended into the previous OFDM symbol and a cyclic suffix is extended into the next OFDM symbol, and both operations cause self-interference. The effect of this self-interference on multipath immunity was explored in \cite{Huang2001} and \cite{Yumin2002} which propose ``orthogonality restoration" to remove the self-interference by cancellation and improve system performance. In order for this cancellation to work, the windowing term is included when reconstructing the interference, but the windowing function is assumed to be known in advance and the authors do not provide a means to estimate it. In this work, we show that this problem can be mitigated without any foreknowledge of the windowing function by way of estimating the window over a set of three adjacent OFDM symbols. Despite uncertainty in the estimate of the window, we show that this signal parameter can offer significant SIC performance improvements. The goal is to estimate a sufficient number of signal parameters to create a copy of the OFDM with enough accuracy to provide significant suppression of the OFDM signal. We use OFDM symbols from IEEE 802.11 as a working example without loss of generality. 

The novel contributions of this paper are as follows:
\begin{itemize}
  \item The work in this paper shows that OFDM windowing has a significant effect on cancellation. It is shown experimentally that including the window results in a 5.3 dB improvement to OFDM signal cancellation over existing methods.

  \item The estimation of OFDM windows is detailed in this paper. We show how to estimate OFDM windows in the presence of other impairments. The accuracy of the estimation in noise is measured. 
  
  \item In this paper, we do not rely on any special symbol alignment, common OFDM numerology, superposition coding as in \cite{Vanka2012}, or some other means implemented to ease the retrieval of lower power users. We rely only on the ability to suppress an OFDM signal through cancellation. 
\end{itemize}

The structure of this paper is as follows. Section \ref{win_est:sec:ofdm_sig_model} describes a signal model for cancellation similar to that in \cite{Clerckx2020RSMA1} using estimates of the channel coefficients and the CFO, without accounting for windowing. Section \ref{win_est:sec:ofdm_windowing} develops a signal model for OFDM with windowing. Section \ref{win_est:sec:win_est} develops an algorithm to estimate the window that includes the presence of a multipath channel and the performance is evaluated. Section \ref{win_est:sec:add_win_to_cancel} applies the algorithms from in sections \ref{win_est:sec:ofdm_sig_model} and \ref{win_est:sec:win_est} to synthetic OFDM packets and shows that estimating the window provides significant improvement in cancellation. Section \ref{win_est:sec:OTA} describes an experiment applying the algorithms in sections \ref{win_est:sec:ofdm_sig_model} and \ref{win_est:sec:win_est} to cancel OFDM signal data collected over-the-air. It is shown that the new method provides a 5.3 dB improvement over existing methods. Section \ref{win_est:sec:conclusion} summarizes the results. Regarding notation, bold upper case, bold lower case, and non-bold lower case letters correspond to matrices, vectors, and scalars, respectively.

\section{OFDM Signal Model without Windowing}
\label{win_est:sec:ofdm_sig_model}
An OFDM symbol is created using an $M$-length Inverse Fast Fourier Transform (IFFT) where $M$ represents the total number of subcarriers, including all pilots and nulls. An OFDM packet is a concatenation of $N$ OFDM symbols in time, where each individual symbol is denoted $o_{p}[ k ]$ as represented in \eqref{eq:packetdefinition}. The subscript ${p}$ indicates the placement of the $p^{th}$ OFDM symbol in the packet and $k$ indexes the individual samples which comprise that OFDM symbol. A single OFDM symbol is length $K$ and is at least $M+L$ long where $L$ is the mandatory cyclic prefix length and $L<M$. As an example, an 802.11g OFDM symbol uses a 64-length IFFT ($M$=64) and an OFDM symbol length of 80 ($K$=80) after a cyclic prefix of 16 ($L$=16). The range $-L \leq k \leq -1$ represents the mandatory cyclic prefix and the range $0 \leq k \leq M-1$ represents the original length-$M$ IFFT.  

\begin{equation}
	\label{eq:packetdefinition}
	\mathbf{s}=\{ \mathbf{o}_0 \| \mathbf{o}_1 ... \| \mathbf{o}_{N-1} \}
\end{equation}

\subsection{Channel and Frequency Offset Impairments}
The samples in time of the OFDM packet as seen by the receiver after timing offset correction, $\mathbf{r}$, is shown in \eqref{eq:rcvdvec} where matrix $\mathbf{H}$ represents the multipath channel, the diagonal matrix $\mathbf{\Lambda_{\Theta}}$ represents the carrier frequency offset, and $\mathbf{n}$ represents noise at the receiver. The carrier offset diagonal matrix $\mathbf{\Lambda_{\Theta}}$ contains phase offsets for each sample of $\mathbf{s}$ as shown in \eqref{eq:lambdatheta}. The magnitude of each element of $\mathbf{\Lambda_{\Theta}}$ is unity. If the carrier frequency offset is constant then when integrated in time it will produce a phase ramp in $\mathbf{\Lambda_{\Theta}}$. The $k^{th}$ element on the diagonal represents a phase offset value $\Theta[k]$ shown as a phase ramp in \eqref{eq:thetak} where  $\omega$ is the frequency offset in radians per sample and $\phi$ is the starting phase offset in radians. 

\begin{equation}
	\mathbf{r}=\mathbf{\Lambda_{\Theta}} \mathbf{H}\mathbf{s}+\mathbf{n}
	\label{eq:rcvdvec}
\end{equation} 

\begin{equation}
	\mathbf{\Lambda_{\Theta}}=\text{diag}(e^{j\mathbf{\Theta}})
	\label{eq:lambdatheta}
\end{equation}

\begin{equation}
	\Theta[k]=\omega k + \phi
	\label{eq:thetak}
\end{equation}

In order to correct the impairments, the corrections derived from estimates of those impairments must be applied to the received samples. Once the start of the 802.11 packet has been determined, estimates are needed for $\mathbf{\Lambda_{\Theta}}$ and $\mathbf{H}$, those being $\widehat{\mathbf{\Lambda_{\Theta}}}$ and $\widehat{\mathbf{H}}$. .   

The first impairment to be estimated is the CFO impairment defined in \eqref{eq:lambdatheta}. The frequency correction estimate $\widehat{\mathbf{\Lambda_{\Theta}}}^{*}$ is applied to the received samples $\mathbf{r}$ before an estimate of the channel matrix can be calculated. Therefore, this error propagates into the estimation of the channel impulse response. The estimate of the channel impulse response is initially calculated using the known sequence in the 802.11 preamble. This only provides estimates for the 52 non-zero subcarriers. The channel estimate is linearly interpolated over the null-subcarriers in phase and magnitude.  

\subsection{Applying Cancellation}
\label{sec:Apply_Cancellation}

The re-modulation produces an OFDM waveform $\mathbf{\widehat{s}}$ representing the estimate of the OFDM signal at the transmitter with no impairments. Applying the corrections to the received waveform may shape the noise and adversely affect signals from other users such as in the case of PD-NOMA. Therefore, the estimated channel $\widehat{\mathbf{H}}$ and the estimated carrier frequency offset $\widehat{\mathbf{\Lambda_\Theta}}$ are applied to $\widehat{\mathbf{s}}$ as shown in \eqref{eq:ResidueNoInverse}.

\begin{equation}
	\mathbf{u}=\Lambda_\Theta \mathbf{H}\mathbf{s}+\mathbf{n}-\widehat{\mathbf{\Lambda_{\Theta}}}\widehat{\mathbf{H}}\widehat{\mathbf{s}}
	\label{eq:ResidueNoInverse}
\end{equation}

Assuming the demodulation process saw no bit errors, then $\widehat{\mathbf{s}}=\mathbf{s}$ and the residue reduces to \eqref{eq:ResidueNoInversereduced}. Cancellation in the presence of bit errors will be shown in section \ref{win_est:sec:win_est}.

\begin{equation}
	\mathbf{u}=(\mathbf{\Lambda_{\Theta}} \mathbf{H}-\widehat{\mathbf{\Lambda_{\Theta}}}\widehat{\mathbf{H}})\mathbf{s}+\mathbf{n}
	\label{eq:ResidueNoInversereduced}
\end{equation}

The signal-error-term of the residue is $(\mathbf{\Lambda_{\Theta}} \mathbf{H}-\widehat{\mathbf{\Lambda_{\Theta}}}\widehat{\mathbf{H}})\mathbf{s}$. As the estimation functions improve, the estimated parameters approach the actual parameters, and $\widehat{\mathbf{\Lambda_{\Theta}}}\widehat{\mathbf{H}}$ approaches $\mathbf{\Lambda_{\Theta}} \mathbf{H}$. As that happens, the signal-error term goes to zero. This would leave only the noise term in the residue. This all assumes the channel, CFO, and noise impairments represent the only significant differences between $\mathbf{r}$ and $\mathbf{s}$. The subsequent sections will show that this is not a safe assumption, that windowing can contribute significantly to this difference, and the lack of a window estimate yields substantially lower cancellation.

\section{OFDM Windowing at the Transmitter}
\label{win_est:sec:ofdm_windowing}
Different windowing schemes employed at the transmitter have been explored in literature, such as \cite{Farhang2011}. Here we analyze the general case. A generalized model for the windowing function $\mathbf{w}$ is provided in \eqref{win_est:eq:windowdef} where subscript $i$ indexes the individual coefficients. Note that the window definition in \eqref{win_est:eq:windowdef} does not assume that the windowing is symmetric. There are two transition regions in the window definition defined by two separate sets of coefficients $\boldsymbol{\alpha}$ and $\boldsymbol{\beta}$. The transition from 1 to 0 at the end of a symbol, including the cyclic suffix, is represented by the coefficients $\boldsymbol{\alpha}$. The transition from 1 to 0 in the cyclic prefix, including any extension thereof, is represented by the coefficients $\boldsymbol{\beta}$.  

The window definition in \eqref{win_est:eq:windowdef} does set some practical limits on the window, without loss of generality. The window is only given nonzero values for  $-2L \leq i \leq M+L-1$. The window values are $1$ for $0 \leq  i \leq M-L-1$. The indices for $\boldsymbol{\alpha}$ and $\boldsymbol{\beta}$ range over $0 \leq i \leq 2L-1$. The length of the $\boldsymbol{\alpha}$ and $\boldsymbol{\beta}$ coefficients is set to $2L$; however, some of these coefficients may be zero or one.    

\begin{equation} 
\label{win_est:eq:windowdef}
w_{i}=
\begin{cases}
0 & \text{for } i > M+L-1 \\
\alpha_{i-M+L} & \text{for } M-L \leq  i \leq M+L-1 \\	
1 & \text{for } 0 \leq  k \leq M-L-1  \\
\beta_{-1-L-i} & \text{for } -2L \leq  i \leq -1 \\
0 & \text{for } i < -2L \\
\end{cases}
\end{equation}

The definition in \eqref{win_est:eq:windowdef} limits the nonzero values in a window applied to any one OFDM symbol to a range $-2L \leq  i \leq M-L-1$. This imposes a maximum length of $M+3L$ on the extended OFDM symbol where $M+L$ represents the standard length of the OFDM symbol, $L$ represents the maximum extension of the cyclic prefix, and the final $L$ term represents the maximum length of the cyclic suffix. The index into $\boldsymbol{\alpha}$ is $i-M+L$ where the offset $-M+L$ represents the start of that transition. The index into $\boldsymbol{\beta}$ is $-1-L-i$ where the term $-i$ reverses the order of the coefficients and where the offset $-1-L$ represents the optional extension of the cyclic prefix.  

\begin{figure}[b!]
	\begin{center}
		\includegraphics[width=\columnwidth]{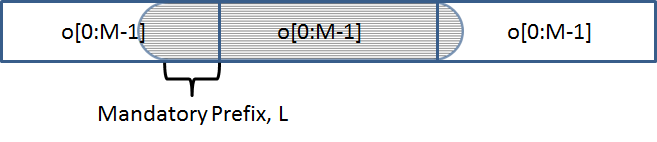}
	\end{center}
	\caption{Cyclic Prefix and Suffix as Repeating Symbols and Window}
	\label{win_est:fig:RepeatPrefixSuffix}
\end{figure}

Windowing is illustrated in Fig. \ref{win_est:fig:RepeatPrefixSuffix}. The figure shows three copies of an OFDM symbol $o[k]$ each of $M$ samples. The repetitions on either end form the cyclic prefix and suffix. The cyclic prefix is longer than the suffix because the cyclic prefix has a mandatory minimum length of $L$. The cyclic suffix and extended cyclic prefix must protrude into the adjacent OFDM symbol. 

The IEEE 802.11 standard contains a windowing recommendation that defines a window transition period spanning equally across the boundary between two OFDM symbols adjacent in time. One half of that transition window extends the cyclic prefix past 16 samples and into the previous OFDM symbol. The other half of the transition window creates a cyclic suffix that extends into the cyclic prefix of the next OFDM symbol.

The windowed OFDM symbols are defined in \eqref{win_est:eq:WindowSigModel}. The sample index $k$ is relative to the current OFDM symbol $\mathbf{o_{p}}$. This is why the previous and next OFDM symbols, $\mathbf{o_{\left( p-1 \right ) \bmod N}}$ and $\mathbf{o_{\left( p+1 \right ) \bmod N}}$ require offsets in the sample index. The window coefficients used on an OFDM symbol are selected by the sample index plus offset applied to that OFDM symbol. Three OFDM symbols are windowed and then summed to create a combined symbol $\mathbf{v_{p}}$. With no windowing $\mathbf{v_{-1}}$ and $\mathbf{v_{1}}$ are always zero, and thus $\mathbf{v}=\mathbf{s}$ where $\mathbf{s}$ is defined in \eqref{eq:packetdefinition}. This process introduces self-interference from $\mathbf{o_{\left( p-1 \right ) \bmod N}}$ and $\mathbf{o_{\left( p+1 \right ) \bmod N}}$ into $\mathbf{o_{p}}$. The summation in \eqref{win_est:eq:WindowSigModel} can be compressed into \eqref{win_est:eq:WindowSigModel2} where $q$ is an integer $ -1 \leq q \leq 1$ representing the previous, current, and next OFDM symbol.   

\begin{equation} 
\label{win_est:eq:WindowSigModel}
\begin{aligned}
&v_{p}[k]= \\ 
& w_{ k+M+L } o_{\left( p-1 \right ) \bmod N} \left [ k+M+L \right ] \\
& + w_{ k } o_{p} \left [ k \right ] \\
& + w_{k-M-L} o_{\left( p+1 \right ) \bmod N} \left [ k-M-L \right ]
\end{aligned}
\end{equation}

\begin{equation} 
\label{win_est:eq:WindowSigModel2}
\begin{aligned}
&v[k]= \sum_{q=-1}^{1} \{ w \left [  k-q \left (M+L \right )  \right ] \\
&o_{(p+q) \bmod N} \left [ k-q \left (M+L \right )  \right ] \}
\end{aligned}
\end{equation}

\begin{figure}[t!]
	\begin{center}
		\includegraphics[width=\columnwidth]{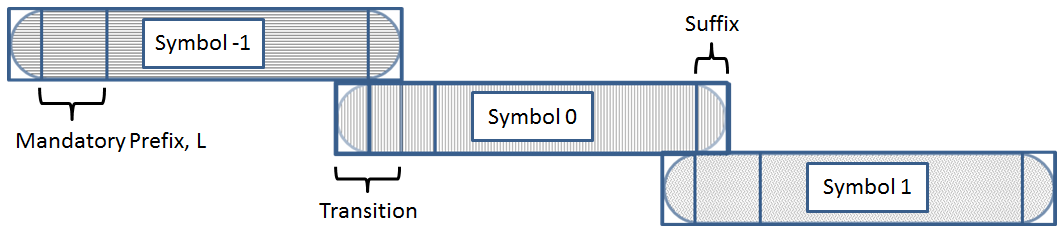}
	\end{center}
	\caption{Overlaying the Extended OFDM Symbols}
	\label{win_est:fig:PrefixSuffix6}
\end{figure}

Fig. \ref{win_est:fig:PrefixSuffix6} illustrates the creation of $\mathbf{v_{p}}$  and the resulting self-interference. The cyclic prefix of $\mathbf{o_{p+1}}$ extends into the end of $\mathbf{o_{p}}$. The cyclic suffix of $\mathbf{o_{p-1}}$ extends into the cyclic prefix of $\mathbf{o_{p}}$. Therefore OFDM symbol $\mathbf{o_{p}}$ has interference from both adjacent symbols. 

The self-interference of windowed OFDM affects multipath immunity. A multipath channel imposes a linear convolutional impairment on the signal. The mandatory cyclic prefix is set to a length sufficient to represent the maximum delay of the multipath channels of the intended environment. In the case of windowed OFDM, the addition of the cyclic suffix from an adjacent OFDM symbol causes interference in the equalization of the current OFDM symbol.

\section{Estimating the OFDM Window}
\label{win_est:sec:win_est}
While using a cyclic suffix and windowing an OFDM symbol is well documented, window estimation for OFDM signals is rarely found in the literature. Estimating the window function is not necessary for demodulation. A cyclic suffix and window are not required in many OFDM standards, such as IEEE 802.11.

A preamble is included at the start of many OFDM packets in order to facilitate estimation of various parameters such as carrier frequency offset and symbol timing phase. That estimate is used to synchronize the received signal. In addition to synchronization impairments, the windowed OFDM signal defined in \eqref{win_est:eq:WindowSigModel} passes through a wireless channel modeled as a linear convolutional impairment shown in \eqref{win_est:eq:WindowAndChannelSigModel}. The term $h$ represents the channel impulse response. An estimate of the channel impulse response can be obtained from the aforementioned preamble. The channel impulse response is estimated in the frequency domain and provides estimates for non-zero subcarriers only. In the case of IEEE 802.11g this represents 52 out of 64 possible subcarriers. The channel estimate is linearly interpolated over the null-subcarriers in phase and magnitude as described in \cite{Shahriar2015}. The channel estimate is then converted to the time domain. The sample index $k-\ell$ in \eqref{win_est:eq:WindowAndChannelSigModel} can take on a value less than $-L$, as allowed by \eqref{win_est:eq:WindowSigModel} for any combined symbol $\mathbf{v_{p}}$.

\begin{equation} 
\label{win_est:eq:WindowAndChannelSigModel}
\begin{multlined}
y_{p}[k]=\sum_{\ell=0}^{L-1}h[\ell]v_{p}[k- \ell] 
\end{multlined}
\end{equation}

The receiver generates an estimate of the received signal $\hat{y}_{p}[k]$ which includes the multipath channel distortion. This estimate is constructed from estimates of the individual OFDM symbols $\hat{o}_{p}$ and an estimate of the multipath channel $\hat{h}$. A squared-error term is defined in \eqref{win_est:eq:errorsqdef} using the estimate $\hat{y}_{p}[k]$. Minimizing the error defined in \eqref{win_est:eq:errorsqdef} with respect to the window coefficients provides an estimate of the window $\hat{w}_{i}[Kp+k]$ where $K$ is the length of an OFDM symbol without windowing. Note that the sample index $k$ is relative to the current symbol. For each OFDM symbol $p$, the sample index $k$ ranges over $-L \leq  k \leq M-1$ thus creating $M+L$ error samples for each OFDM symbol used. As an example, for an 802.11 OFDM sample, there would be 80 (64+16) error samples produced per OFDM symbol. The estimation of the window is performed over $N$ OFDM symbols each of length $K$ samples.

\begin{equation} 
\label{win_est:eq:errorsqdef}
|e[Kp+k]|^2=\left ( y_{p}[k] - \hat{y}_{p}[k]   \right ) \left ( y_{p}^*[k] - \hat{y}_{p}^*[k]   \right )
\end{equation}

Taking the derivative of \eqref{win_est:eq:errorsqdef} with respect to the estimate of the window $\hat{w}_{i}[Kp+k]$ where $i = k -q \left (M+L \right ) - \ell$, and substituting $v_{p}[k- \ell]$ with \eqref{win_est:eq:WindowSigModel2}, we find the derivative shown in \eqref{win_est:eq:windowupdate}. The derivative in \eqref{win_est:eq:windowupdate} reduces to \eqref{win_est:eq:windowupdatesimplified}.

\begin{equation} 
\label{win_est:eq:windowupdate}
\begin{aligned}
&\frac{\partial |e[Kp+k]|^2}
{
	\partial \hat{w}_{i}[Kp+k]
} = \\
&-e[Kp+k] \hat{h}^{*} [\ell]
\hat{o}^{*}_{(p+q) \bmod N} [i] \\
&-e^{*}[Kp+k] \hat{h} [\ell]
\hat{o}_{(p+q) \bmod N} [i]
\end{aligned}
\end{equation}


\begin{equation} 
\label{win_est:eq:windowupdatesimplified}
\begin{aligned}
&\frac{\partial |e[Kp+k]|^2}
{
	\partial \hat{w}_{i}[Kp+k]
} =\\
&-2\Re \{  
e^{*}[Kp+k] \hat{h} [\ell]
\hat{o}_{(p+q) \bmod N} [i] 
\} 
\end{aligned}
\end{equation}

The update step is defined in \eqref{win_est:eq:windowupdatemath}. For every OFDM symbol there are $L$ updates to each window coefficient estimate as the channel $h$ is $L$ coefficients long. Each OFDM symbol is indexed by $p$ indicating the sequential placement of that OFDM symbol in the OFDM packet. The update function in \eqref{win_est:eq:windowupdatesimplified} is scaled by a step size $\mu$ and accumulated into estimates of all $\hat{w}_{ i }[Kp+k+1]$ for each valid index $i$. The values for $\boldsymbol{\alpha}$ are the window coefficient estimates $\hat{w}_{ i }[Kp+k+1]$ where $M-L \leq  i  \leq M+L-1$. The values for $\boldsymbol{\beta}$ are the window coefficient estimates $\hat{w}_{ i }[Kp+k+1]$ where $-2L \leq  i  \leq -1$.    

\begin{equation} 
\label{win_est:eq:windowupdatemath}
\begin{aligned}
&\hat{w}_{i}[Kp+k+1]= \hat{w}_{i}[Kp+k] \\
&+ 2 \mu \Re \{  
e^{*}[Kp+k] \hat{h} [\ell]
\hat{o}_{(p+q) \bmod N} [i] 
\} 
\end{aligned}
\end{equation}




The process of estimating an OFDM window is as follows: when receiving an OFDM packet, demodulate all symbols. Use the demodulated bits to create a new OFDM packet. The received OFDM packet is $y_{p}[k]$ and the locally generated one is $\hat{y}_{p}[k]$. Use no windowing on $\hat{y}_{p}[k]$. The estimate $\hat{y}_{p}[k]$ is used as the reference in \eqref{win_est:eq:errorsqdef} to which the received data measurement $y_{p}[k]$ is compared. The received OFDM signal $y_{p}[k]$ and the estimate $\hat{y}_{p}[k]$ are then used as per \eqref{win_est:eq:windowupdatesimplified} to create an estimate of the window $\hat{w}$.

\begin{figure}[b!]
	\begin{center}
		\includegraphics[width=\columnwidth]{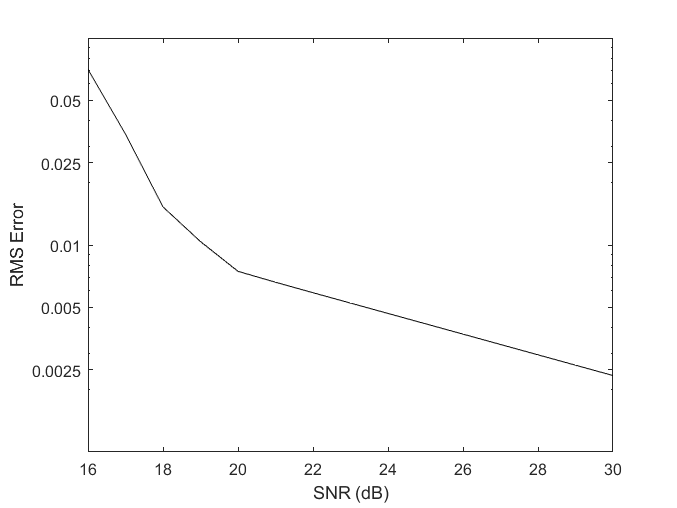}
	\end{center}
	\caption{Window Error as a function of SNR}
	\label{win_est:fig:WinErrorSim}
\end{figure}

Adding noise into the window estimation creates uncertainty not only in the measurement $y_{p}[k]$ but also in the reference $\hat{y}_{p}[k]$ to which that measurement is compared because the added noise creates bit errors. Those bit errors cause errors in the estimate $\hat{y}_{p}[k]$ and that reduces the total accuracy of the window estimation. Fig. \ref{win_est:fig:WinErrorSim} shows the average RMS error resulting from estimating the window of a received OFDM packet as a function of SNR. The estimate is performed on an IEEE 802.11 packet using 64 QAM. The channel estimate is perfect for this measurement. Each data point in Fig. \ref{win_est:fig:WinErrorSim} represents an estimate performed over 148 OFDM symbols using a step size of 0.01 and 20 epochs. The window defined in \cite{80211Standard2020} and the parameter \emph{transition time} serves as a roll-off. Here the transition time is set to 500 ns for ease of visualization. The window error is not linear with SNR because the frequency of bit errors increases as SNR decreases, adding additional uncertainty into the estimate. 

\begin{figure}[b!]
	\begin{center}
		\includegraphics[width=\columnwidth]{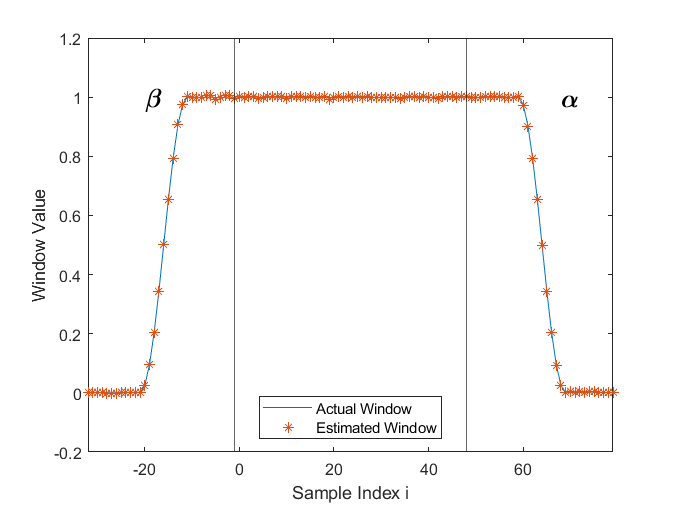}
	\end{center}
	\caption{Actual and Estimated Window Overlaid}
	\label{win_est:fig:ActualEstim}
\end{figure}

Fig. \ref{win_est:fig:ActualEstim} shows the actual and estimated window overlaid. The window was estimated with an SNR of 30 dB using 20 epochs. The index $i$ ranges from $-32 \leq i \leq 79$, that being $-2L \leq i \leq M+L-1$. The regions representing $\boldsymbol{\alpha}$ and $\boldsymbol{\beta}$ are shown.

\section{Cancelling OFDM Signals with Imperfect Window and Channel Estimates}
\label{win_est:sec:add_win_to_cancel}
Combining the definition of windowing in \eqref{win_est:eq:WindowSigModel2} with that of cancellation in \eqref{eq:ResidueNoInverse} results in \eqref{eq:cancelwithwindow}. For the purposes of cancellation, windowing is represented as a diagonal matrix $\mathbf{\Lambda_{w_{q}}}$ estimated as $\widehat{\mathbf{\Lambda_{w_{q}}}}$. Each element is $\mathbf{\Lambda_{w_{q}}}$ is the window value applied to the signal $\mathbf{s_{q}}$ offset by $q$ as in \eqref{win_est:eq:WindowSigModel2}. Three copies of the signal $\mathbf{s}$ without windowing are created, offset in samples, and then windowed by the respective $\mathbf{\Lambda_{w_{q}}}$. The channel and CFO estimates will be imperfect as in \eqref{eq:ResidueNoInverse}.

\begin{equation}
	\mathbf{u}=\mathbf{\Lambda_{\Theta}} \mathbf{H}\sum_{q=-1}^{1}\mathbf{\Lambda_{w_{q}}}\mathbf{s_{q}}+\mathbf{n}-\widehat{\mathbf{\Lambda_{\Theta}}}\widehat{\mathbf{H}}\sum_{q=-1}^{1}\widehat{\mathbf{\Lambda_{w_{q}}}}\widehat{\mathbf{s_{q}}}
	\label{eq:cancelwithwindow}
\end{equation}

\begin{figure}[b!]
	\begin{center}
		\includegraphics[width=\columnwidth]{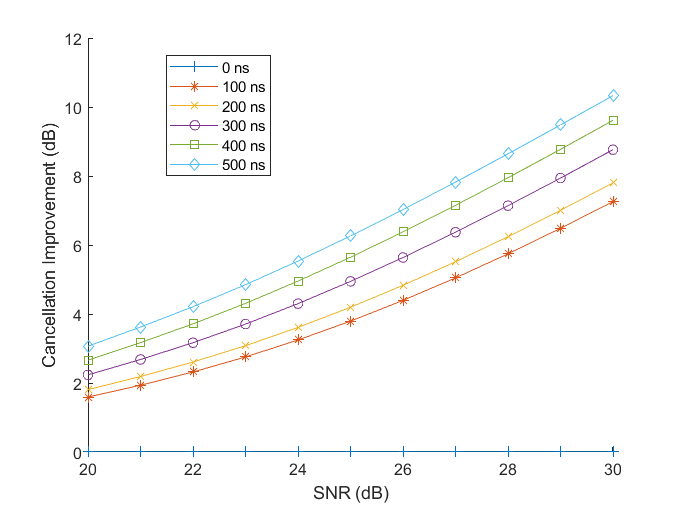}
	\end{center}
	\caption{Cancellation Improvement as a function of SNR}
	\label{win_est:fig:CancelImprove}
\end{figure}

\begin{figure}[t!]
	\begin{center}
		\includegraphics[width=\columnwidth]{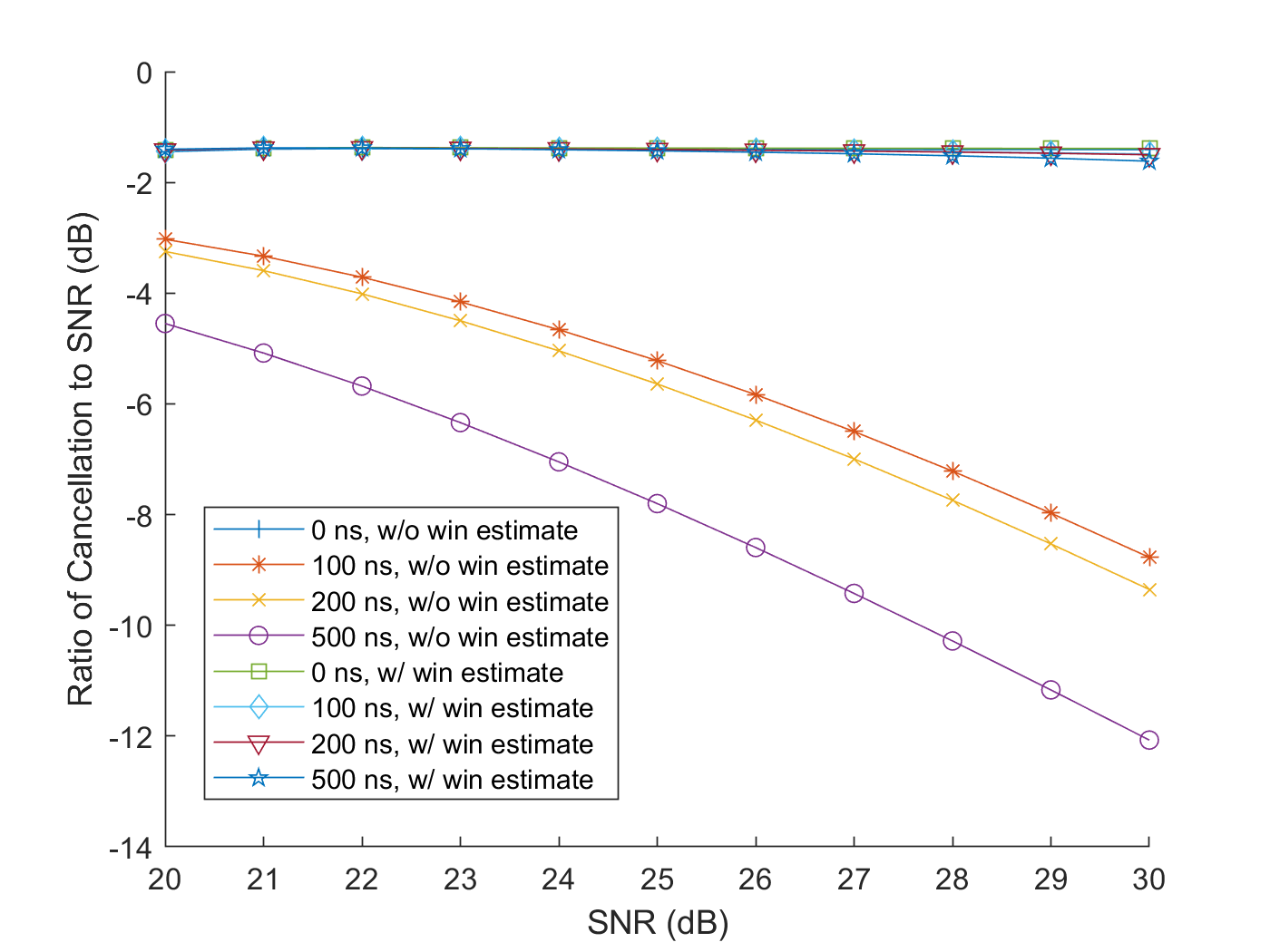}
	\end{center}
	\caption{Ratio of Cancellation to SNR as a function of SNR}
	\label{win_est:fig:CancelToSnr}
\end{figure}

In this experiment, we generate synthetic 500 OFDM packets each with a payload of 4000 bytes. We apply raised cosine windows with a range of transition times to the packets from 100 ns to 1 $\mu$s. A window transition time of 100 ns represents the smallest window that can be applied to the 802.11 packet. We execute the window estimation algorithm across a range of SNR from 20 to 30 dB. We then perform cancellation with and without the window estimate following \eqref{eq:ResidueNoInverse} and \eqref{eq:cancelwithwindow}. The cancellation $c$ is calculated as $c=var(\mathbf{r})/var(\mathbf{u})$ that being the ratio of the variance of the received samples $\mathbf{r}$ to that of the residue $\mathbf{u}$. The ratio of the cancellation $c$ with windowing to without windowing for six window transition values are plotted in Fig. \ref{win_est:fig:CancelImprove}. The two methods are equal (0 dB) when no window is present. As soon as even a small window is applied, the cancellation method with windowing falls well short of that without. In Fig. \ref{win_est:fig:CancelToSnr} the power of the residue resulting from each test case is subtracted from the SNR of that test case. The SNR represents the absolute maximum cancellation, as then the residue would be noise. The results are all negative values demonstrating all test cases fall short of perfect estimates. The test cases employing windowing estimation keep much closer to a ratio 0 dB ratio.



\section{OTA Cancellation Residue Reduction}
\label{win_est:sec:OTA}
\begin{figure}[b!]
	\begin{center}
		\includegraphics[width=\columnwidth]{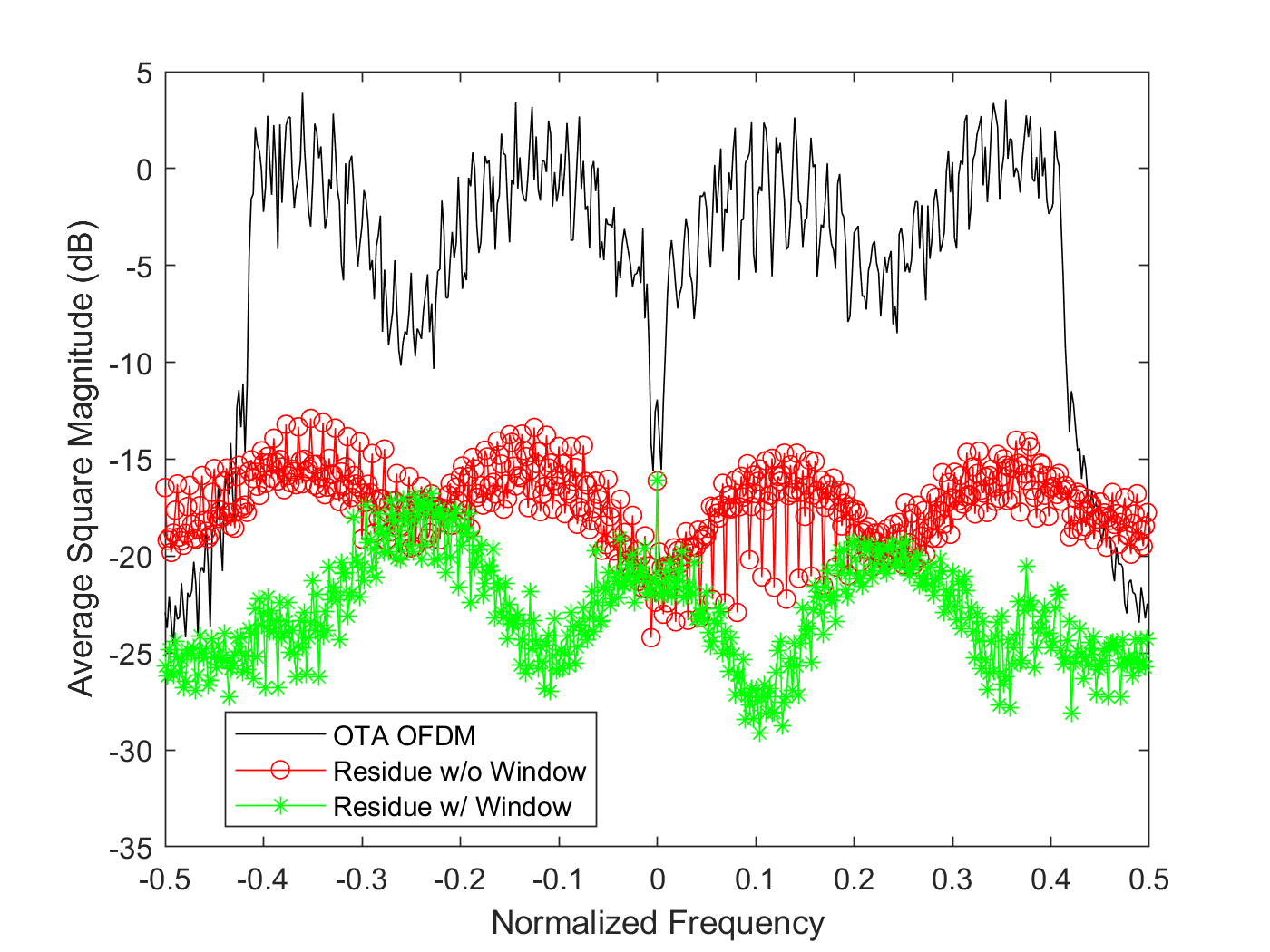}
	\end{center}
	\caption{Spectrum of OTA Packet and Two Residues}
	\label{win_est:fig:residuespectrums}
\end{figure}

In this experiment, the two cancellation algorithms were applied to 961 OFDM packets captured OTA. The SNR estimates of the OTA packets varied from 29 to 31 dB. The modulation scheme used by all captured OFDM packets for the data was QAM64. The average cancellation without estimating the window as in \eqref{eq:ResidueNoInverse} was 19.5 dB. The average cancellation with estimating windowing as in \eqref{eq:cancelwithwindow} was 5.3 dB higher. Both the base cancellation without windowing and the windowing improvement fell short of expectations from the experiment with synthetic data, however, this experiment demonstrates unequivocally that the window estimation provided a substantial boost in cancellation. Fig. \ref{win_est:fig:residuespectrums} is a spectrum plot showing a single recorded OTA packet and the residue from the cancellation algorithms with and without windowing. This figure illustrates the impact of including windowing in the cancellation algorithm. 

\section{Conclusion}
\label{win_est:sec:conclusion}
We have presented and evaluated the performance of both an OFDM window estimation method and a technique to use that estimate to cancel OFDM signals. We presented the window estimation method in a generalized form that does not require foreknowledge of the window implementation. We use 802.11 as an example application for this method. We show that the algorithm works in the presence of noise and imperfect channel estimates, and provides a significant boost to OFDM signal cancellation with synthetic data. We demonstrate that the algorithm provides a boost when using OTA data. Windowing is a common method used to limit the bandwidth of OFDM signals, and an estimator for the window should be present in an algorithm to cancel an OFDM signal. 

\bibliographystyle{IEEEtran}
\bibliography{IEEEabrv,WinEstimation}

\end{document}